\documentclass[11pt]{article}
\usepackage{amssymb}
\usepackage{amsfonts}
 \usepackage{pifont}
\usepackage{amsmath,amsfonts,amssymb,theorem,graphicx,listings,txfonts}
\usepackage{graphics}
 \usepackage{caption2}
\usepackage{flafter}
 \usepackage{indentfirst}
\usepackage{epsfig}
  \usepackage{mathrsfs}
\usepackage{subfigure}
\usepackage{wrapfig}
 \usepackage{cite}
 \usepackage{morefloats}
\newtheorem{theorem}{Theorem}[section]

\newtheorem{definition}[theorem]{Definition}

\theoremstyle{example}
\newtheorem{example}[theorem]{Example}
\theoremstyle{procedure}

\theoremstyle{experiment}

\theoremstyle{problem}

\renewcommand{\arraystretch}{1.5}

\title{Decision-theoretic rough sets based on time-dependent loss function}
\author
{Guangming Lang$^{a}$\thanks{Corresponding author.\quad Tel./fax:
+86 731 88822855, langguangming1984@126.com(G.M. Lang) \hspace{1cm}
\newline\mbox{}\hspace{0.55cm}
E-mail address: cmjlong@163.com(M.J. Cai). } \hspace{1cm} Mingjie
Cai$^{b}$\\
\small {$^{a}$ School of Mathematics and Computer Science, Changsha University of Science and Technology}\\
\small {Changsha, Hunan 410082, P.R. China}\\
\small {$^{b}$ College of Mathematics and Econometrics, Hunan University}\\
\small {Changsha, Hunan 410082, P.R. China} }
\date{}

\begin{document}
\maketitle \baselineskip=17pt
\begin{center}
\begin{quote}
{{\bf Abstract.} A fundamental notion of decision-theoretic rough
sets is the concept of loss functions, which provides a powerful tool
of calculating a pair of thresholds for making a decision with a minimum cost. In this paper, time-dependent loss functions which
are variations of the time are of interest because such functions
are frequently encountered in practical situations,
we present the relationship between the pair of thresholds and
loss functions satisfying time-dependent uniform distributions and normal
processes in light of bayesian decision procedure. Subsequently, with the aid of
bayesian decision procedure, we
provide the relationship between the pair of thresholds and loss
functions which are time-dependent interval sets and fuzzy numbers. Finally, we employ several
examples to illustrate that how to calculate the thresholds for making a
decision by using time-dependent loss functions-based decision-theoretic
rough sets.

\\}
\end{quote}
\end{center}
\renewcommand{\thesection}{\arabic{section}}

\section{Introduction}

Rough set theory, proposed by Pawlak\cite{Pawlak1} in 1982, is a
powerful mathematical tool to deal with uncertainty, imprecise or
incomplete knowledge for information systems. But the condition of
the equivalence relation in Pawlak's model is so strict that limits
its applications. To generalize Pawlak's rough sets, researchers
have presented various kinds of probabilistic rough sets
(PRS) such as decision-theoretic rough sets
(DTRS)\cite{Hu1,Jia1,Yao2,Yao5,Yao4,Yao10,Zhang4,
Liu11,Slezak1},
bayesian rough sets (BRS)\cite{Ma1,Yao8} and
game-theoretic rough sets (GTRS)\cite{Azam1,Azam2} for solving
practical problems. To date, probabilistic rough set models have
been successfully applied to many fields such as data mining, email
spam filtering, investment management and web support.

Since PRS was proposed, the determination of a pair of thresholds has become a substantial challenge. Until now, researchers have
presented some reasonable semantic interpretations for the pair of
thresholds. For example, Cheng et
al.\cite{Cheng1} computed precision parameter values based on
inclusion degree with variable precision rough set model. Deng and Yao\cite{Deng1,Deng2} presented an
information-theoretic approach to the interpreation and
determination of thresholds used in PRS and presented DTRS-based three-way approximations of fuzzy sets.
Herbert and JT Yao\cite{Herbert1,Herbert2} proposed GTRS to determine
the values of thresholds used in PRS by introducing game theory and
investigated its capability of analyzing a major decision problem
evident in existing PRS.
Yao\cite{Yao4} proposed DTRS, which provides a new interpretation in
the aspect of determining the threshold vlaues by using loss
functions, by combining bayesian decision theory with PRS. To trade
off different types of classification error, three-way
decision-theory was proposed by Yao for making a decision with the
minimum cost on the basis of DTRS, whereas there are three choices
of acceptance, deferment and rejection. Concretely, rules from the
positive region are used for making a decision of acceptance, rules
from the negative region are applied to make a decision of
rejection, and rules from the boundary region are used for making a
decision of deferment. More specially, the choice deferment reduces
the loss of making a decision in DTRS. Therefore, DTRS provides a
powerful tool for making a decision with a minimum cost ternary
classifier.

In DTRS, three choices of acceptance, deferment
and rejection are determined by loss functions. In recent years,
many investigations have been done on loss functions for DTRS in literatures. For example, Jia et al.\cite{Jia1,Jia2} conducted the minimum cost attribute reduction
in decision-theoretic rough set models and presented an optimization
representation of decision-theoretic rough set model.
Li and Zhou\cite{Li2} gave two assumptions for the values of losses and proposed a multi-view DTRS decision model. Liu, Li and Liang\cite{Liu6} performed three-way government
decision analysis with decision-theoretic rough sets. Liu, Yao and Li\cite{Liu8} proposed a profit-based three-way approach to the investment decision-making and utilized the objects to estimate the losses or carry out some questionnaires or behavioral experiments. Liu, Li and Ruan\cite{Liu10} investigated probabilistic model criteria with decision-theoretic rough sets.  Yao\cite{Yao4} used relative values between losses to express the thresholds and reduced the variable amount of the thresholds. Furthermore, Liang et al.\cite{Liang1}
presented triangular fuzzy
decision-theoretic rough sets by considering bayesian decision procedure, in which loss functions are triangular fuzzy numbers. Liang and Liu\cite{Liang2}
provided systematic studies on three-way decisions with interval-valued
decision-theoretic rough sets, in which loss functions are interval-valued. Liu, Li and Liang\cite{Liu9} proposed dynamic decision-theoretic rough sets, in which loss functions are single-valued variations of time.
Up to now DTRS
has been successfully applied to expert system, medical diagnosis,
environmental science, conflict analysis and economics. Accordingly,
applications are increasingly being adopted with the development of
DTRS.

In practical situations, time-dependent loss functions are of interest because such functions are frequently encountered. For example, if we intend to make omelets for breakfast
with six eggs and have cracked five eggs into a bowl, then we will
crack the sixth egg into the bowl. There are two situations for the
sixth egg: bad and good. If we crack a good egg into the bowl, then
an omelet with six eggs is prepared for breakfast. If we crack a bad
egg into the bowl, then five eggs will be lost. If the price of an
egg is 1 unit now, then the loss is five units. If the price of an
egg is 1.2 unit tomorrow, then the loss is 6 units. Clearly, the
loss is the variation of the time since the price of an egg is
varying with the time. If the bad egg is cracked into another bowl,
then the loss is to wash one more bowl, and the expense of washing a
bowel is also varying with the time.
Furthermore, loss functions are not only variations of time but also satisfied some distributions such as uniform distributions, normal processes, interval sets and
fuzzy numbers simultaneously. Therefore, it is urgent to further
study time-dependent loss functions for making a decision by using
three-way decision-theory.

The purpose of this paper is to further investigate time-dependent loss functions-based DTRS. Section 2 introduces the
basic principles of DTRS. Section 3
calculates the values of thresholds when loss functions are satisfied time-dependent uniform distributions and normal processes.
Section 4 is devoting to
studying the relationship between the values of thresholds and loss functions which are time-dependent interval sets.
Section 5 presents the relationship between the values of thresholds and loss functions which are time-dependent fuzzy numbers. The
conclusion comes in Section 6.

\section{Current research on DTRS}

In this section, we review some concepts of decision-theoretic rough sets.

Suppose $S=(U,A,V,f)$ is an information system, $\forall X\subseteq U$
and $0\leq \beta \leq\alpha\leq 1$, the $(\alpha,\beta)-$
probabilistic lower and upper approximations of $X$ are defined as
follows:
\begin{eqnarray*}
\underline{apr}_{(\alpha,\beta)}(X)=\{x\in U: P(X|[x])\geq
\alpha\};
\overline{apr}_{(\alpha,\beta)}(X)=\{x\in U: P(X|[x])> \beta\},
\end{eqnarray*} where $P(X|[x]=\frac{|[x]\cap X|}{|[x]|}$ is the conditional probability of an object $x$
belonging to $X$ when the object is described by its equivalence
class $[x]$.
On the basis of $(\alpha,\beta)-$
probabilistic lower and upper approximation operators, we have the $(\alpha,\beta)-$
probabilistic positive, boundary and negative regions as
follows:
\begin{eqnarray*}
POS_{(\alpha,\beta)}(X)&=&\underline{apr}_{(\alpha,\beta)}(X)=\{x\in U: P(X|[x])\geq \alpha\};
\\
BND_{(\alpha,\beta)}(X)&=&\overline{apr}_{(\alpha,\beta)}(X)-\underline{apr}_{(\alpha,\beta)}(X)=\{x\in U: \beta< P(X|[x])<\alpha\};\\
NEG_{(\alpha,\beta)}(X)&=&U-\overline{apr}_{(\alpha,\beta)}(X)=\{x\in U: P(X|[x])\leq \beta\}.
\end{eqnarray*}

In DTRS, rules from the positive region are used for making a
decision of acceptance, rules from the negative region are used for making a decision of rejection, and rules from the boundary region
are used for making a decision of deferment.
In practical situations, it is hard to acquire the values of the
parameters $\alpha$ and $\beta$ since they are subjective.

To determine the pair of thresholds objectively, Yao\cite{Yao4} proposed decision-theoretic rough sets by
combining bayesian decision theory with PRS.
Concretely, decision-theoretic rough sets model contains $2$ states
$(\Omega=\{X,\neg X\})$ and $3$ actions
$(\mathscr{A}=\{a_{P},a_{B},a_{N}\})$, where $X $ and $\neg X$
indicate that an object is in $X$ and not in $X$, respectively, and
$a_{P},a_{B}$ and $a_{N}$ denote three actions in classifying an
object $x$ into $POS(X)$, $BND(X)$ and $NEG(X)$, respectively. In
Table 1, $\lambda_{PP},\lambda_{BP}$ and $\lambda_{NP}$ denote
losses of taking actions of $a_{P},a_{B}$ and $a_{N}$, respectively,
when an object belongs to $X$; $\lambda_{PN},\lambda_{BN}$ and
$\lambda_{NN}$ denote losses of taking actions of $a_{P},a_{B}$ and
$a_{N}$, respectively, when an object belongs to $\neg X$.

\begin{table}[htbp]\renewcommand{\arraystretch}{1.5}
\caption{Loss function.}
 \tabcolsep0.9in
\begin{tabular}{c c c }
\hline $Action$  & $X(P)$ &$\neg X(N)$\\
\hline
$a_{P}$ & $\lambda_{PP}$& $\lambda_{PN}$  \\
$a_{B}$ & $\lambda_{BP}$& $\lambda_{BN}$ \\
$a_{N}$ & $\lambda_{NP}$& $\lambda_{NN}$  \\
\hline
\end{tabular}
\end{table}

Suppose $\lambda_{PP}\leq\lambda_{BP}\leq \lambda_{NP}$ and
$\lambda_{NN}\leq\lambda_{BN}\leq\lambda_{PN}$, since
$P(X|[x])+P(\neg X|[x])=1$, the bayesian decision procedure suggests
the following minimum-cost decision rules:

$(P):$ If $P(X|[x])\geq \alpha$, then
$x\in POS(X);$

$(B):$ If $\beta<P(X|[x])< \alpha$, then $x\in
BND(X);$

$(N):$ If $P(X|[x])\leq \beta$, then $x\in
NEG(X),$ where
\begin{eqnarray*}
\alpha=\frac{\lambda_{PN}-\lambda_{BN}}{\lambda_{PN}-\lambda_{BN}+\lambda_{BP}-\lambda_{PP}},
\beta=\frac{\lambda_{BN}-\lambda_{NN}}{\lambda_{BN}-\lambda_{NN}+\lambda_{NP}-\lambda_{BP}}.
\end{eqnarray*}

In practice, loss functions are variations of the time, and it is of interest to study the relationship between the thresholds and time-dependent loss functions.

\section{Time-dependent uniform distributions and normal processes-based
DTRS}

In this section, we
investigate DTRS when loss functions are
satisfied time-dependent uniform distributions and normal processes.

For a random variable $X$, there are two common probability
density functions
\makeatother $$f(x)=\left\{
\begin{array}{ccc}
\frac{1}{b-a},&{\rm if}& a\leq x\leq b ;\\
0,&{\rm }& otherwise.
\end{array}
\right.
\text{ and } \makeatother f(x, \mu,
\sigma^{2})=\frac{1}{\sigma\sqrt{2\pi}}e^{-\frac{(x-\mu)^{2}}{2\sigma^{2}}},$$
Then $X$ are said to be satisfied uniform
distribution on $[a, b]$ and normal process which is satisfied
mathematical expectations $\mu$ and variance $\sigma^{2}$, respectively, denoted as $X\sim U(a, b)$,
and $X\sim U(\mu, \sigma^{2})$, where $a$, $b$, $\mu$ and $\sigma$ are constants. In practice, the probability
density functions are varying with time, and there is a need to study DTRS when loss functions are satisfied time-dependent probability
density functions.

\subsection{Time-dependent uniform distributions-based DTRS}

In this subsection, we introduce the concept of time-dependent uniform
distribution for DTRS.

\begin{definition}
Let $X(t)$ be a variation of the time $t$, and the probability
density function of $X(t)$ is \makeatother $$f(x, t)=\left\{
\begin{array}{ccc}
\frac{1}{b(t)-a(t)},&{\rm if}& a(t)\leq x\leq b(t) ;\\
0,&{\rm }& otherwise.
\end{array}
\right. $$  Then $X(t)$ is said to be satisfied time-dependent uniform
distribution on $[a(t), b(t)]$, denoted as $X(t)\sim U(a(t), b(t))$.
\end{definition}

In what follows, we employ Table 2 to illustrate a time-dependent loss
function, where
$\lambda_{PP}(t),\lambda_{BP}(t),\\\lambda_{NP}(t),\lambda_{PN}(t),\lambda_{BN}(t)$
and $\lambda_{NN}(t)$ are varying with the time. In Table 2,
$\lambda_{PP}(t),\lambda_{BP}(t)$ and $\lambda_{NP}(t)$ denote
losses of taking actions of $a_{P},a_{B}$ and $a_{N}$, respectively,
when an object belongs to $X$; $\lambda_{PN}(t),\lambda_{BN}(t)$ and
$\lambda_{NN}(t)$ denote losses of taking actions of $a_{P},a_{B}$
and $a_{N}$, respectively, when an object belongs to $\neg X$.

Subsequently, we discuss DTRS when loss
function is satisfied time-dependent uniform distributions.

\begin{table}[htbp]\renewcommand{\arraystretch}{1.5}
\caption{Time-dependent loss function.}
 \tabcolsep0.9in
\begin{tabular}{c c c }
\hline $Action$  & $X(P)$ &$\neg X(N)$\\
\hline
$a_{P}$ & $\lambda_{PP}(t)$& $\lambda_{PN}(t)$  \\
$a_{B}$ & $\lambda_{BP}(t)$& $\lambda_{BN}(t)$ \\
$a_{N}$ & $\lambda_{NP}(t)$& $\lambda_{NN}(t)$  \\
\hline
\end{tabular}
\end{table}

\begin{theorem}
Let $\lambda_{PP}(t)\sim U(a_{PP}(t), b_{PP}(t)),
\lambda_{BP}(t)\sim U(a_{BP}(t), b_{BP}(t)), \lambda_{NN}(t)\sim
U(a_{NN}(t), b_{NN}(t)), \lambda_{NP}(t)\\\sim U(a_{NP}(t),
b_{NP}(t)), \lambda_{BN}(t)\sim U(a_{BN}(t), b_{BN}(t))$ and $
\lambda_{PN}(t)\sim U(a_{PN}(t), b_{PN}(t))$, where $0\leq
a_{PP}(t)\leq a_{BP}(t)\leq a_{NP}(t), 0\leq a_{NN}(t)\leq
a_{BN}(t)\leq a_{PN}(t), 0\leq b_{PP}(t)\leq b_{BP}(t)\leq
b_{NP}(t)$, $0\leq b_{NN}(t)\leq b_{BN}(t)\leq b_{PN}(t)$ and $t\in
T$. Then we have the following rules:

$(1)$ If $P(X|[x])\geq \alpha(t)$ ,
then  $x\in POS(X);$

$(2)$ If $\beta(t)<P(X|[x])< \alpha(t)$,  then
$x\in BND(X);$

$(3)$ If $P(X|[x])\leq \beta(t)$, then
$x\in NEG(X),$ where
\begin{eqnarray*}
\alpha(t)&=&\frac{[a_{PN}(t)+b_{PN}(t)]-[a_{BN}(t)+b_{BN}(t)]}{[a_{PN}(t)+b_{PN}(t)]-[a_{BN}(t)+b_{BN}(t)]+[a_{BP}(t)+b_{BP}(t)]-[a_{PP}(t)+b_{PP}(t)]};\\
\beta(t)&=&\frac{[a_{BN}(t)+b_{BN}(t)]-[a_{NN}(t)+b_{NN}(t)]}{[a_{BN}(t)+b_{BN}(t)]-[a_{NN}(t)+b_{NN}(t)]+[a_{NP}(t)+b_{NP}(t)]-[a_{BP}(t)+b_{BP}(t)]}.
\end{eqnarray*}
\end{theorem}

\noindent\textbf{Proof.} Since $\lambda_{PP}(t)\sim U(a_{PP}(t),
b_{PP}(t)), \lambda_{BP}(t)\sim U(a_{BP}(t), b_{BP}(t)),
\lambda_{NN}(t)\sim U(a_{NN}(t), b_{NN}(t)), \lambda_{NP}(t)\sim
U(a_{NP}(t), b_{NP}(t)), \lambda_{BN}(t)\sim U(a_{BN}(t),
b_{BN}(t))$ and $ \lambda_{PN}(t)\sim U(a_{PN}(t), b_{PN}(t))$,
where $0\leq a_{PP}(t)\leq a_{BP}(t)\leq a_{NP}(t), 0\leq
a_{NN}(t)\leq a_{BN}(t)\leq a_{PN}(t), 0\leq b_{PP}(t)\leq
b_{BP}(t)\leq b_{NP}(t)$, $0\leq b_{NN}(t)\leq b_{BN}(t)\leq
b_{PN}(t)$ and $t\in T$, we have
\begin{eqnarray*}
&&\frac{a_{PP}(t)+b_{PP}(t)}{2}\leq \frac{a_{BP}(t)+b_{BP}(t)}{2}
\leq \frac{a_{NP}(t)+b_{NP}(t)}{2},\\
&&\frac{a_{NN}(t)+b_{NN}(t)}{2}\leq \frac{a_{BN}(t)+b_{BN}(t)}{2}
\leq \frac{a_{PN}(t)+b_{PN}(t)}{2}.
\end{eqnarray*}

By taking
\begin{eqnarray*}
\lambda_{PP}(t)&=&\frac{a_{PP}(t)+b_{PP}(t)}{2},
\lambda_{PN}(t)=\frac{a_{NN}(t)+b_{NN}(t)}{2}, \lambda_{BP}(t)=
\frac{a_{BP}(t)+b_{BP}(t)}{2},\\ \lambda_{BN}(t)&=&
\frac{a_{BN}(t)+b_{BN}(t)}{2},
\lambda_{NP}(t)=\frac{a_{NP}(t)+b_{NP}(t)}{2}, \lambda_{NN}(t)=
\frac{a_{PN}(t)+b_{PN}(t)}{2},
\end{eqnarray*}
we have the expected losses $R(a_{P}|[x])$, $R(a_{B}|[x])$ and
$R(a_{N}|[x])$ associated with taking the individual actions for an
object $x$ and $t\in T$ as follows:
\begin{eqnarray*}
R(a_{P}|[x])=\lambda_{PP}(t)P(X|[x])+\lambda_{PN}(t)P(\neg
X|[x]);\\
R(a_{B}|[x])=\lambda_{BP}(t)P(X|[x])+\lambda_{BN}(t)P(\neg
X|[x]);\\
R(a_{N}|[x])=\lambda_{NP}(t)P(X|[x])+\lambda_{NN}(t)P(\neg X|[x]).
\end{eqnarray*}

The bayesian decision procedure suggests the local minimum-cost
decision rules:

$(P):$ If $R(a_{P}|[x])\leq R(a_{B}|[x])$ and $R(a_{P}|[x])\leq
R(a_{N}|[x])$, then $x\in POS(X);$

$(B):$ If $R(a_{B}|[x])\leq R(a_{P}|[x])$ and $R(a_{B}|[x])\leq
R(a_{N}|[x])$, then $x\in BND(X);$

$(N):$ If $R(a_{N}|[x])\leq R(a_{P}|[x])$ and $R(a_{N}|[x])\leq
R(a_{B}|[x])$, then $x\in NEG(X).$

Since $P(X|[x])+P(\neg X|[x])=1$, we have $\alpha(t)$ and $\beta(t)$ as follows:
\begin{eqnarray*}
\alpha(t)&=&\frac{\lambda_{PN}(t)-\lambda_{BN}(t)}{[\lambda_{PN}(t)-\lambda_{BN}(t)]+[\lambda_{BP}(t)-\lambda_{PP}(t)]}\\
&=&\frac{[a_{PN}(t)+b_{PN}(t)]-[a_{BN}(t)+b_{BN}(t)]}{[a_{PN}(t)+b_{PN}(t)]-[a_{BN}(t)+b_{BN}(t)]+[a_{BP}(t)+b_{BP}(t)]-[a_{PP}(t)+b_{PP}(t)]};
\\
\beta(t)&=&\frac{\lambda_{BN}(t)-\lambda_{NN}(t)}{[\lambda_{BN}(t)-\lambda_{NN}(t)]+[\lambda_{NP}(t)-\lambda_{BP}(t)]}\\
&=&\frac{[a_{BN}(t)+b_{BN}(t)]-[a_{NN}(t)+b_{NN}(t)]}{[a_{BN}(t)+b_{BN}(t)]-[a_{NN}(t)+b_{NN}(t)]+[a_{NP}(t)+b_{NP}(t)]-[a_{BP}(t)+b_{BP}(t)]}.
\end{eqnarray*}

On the basis of $\alpha(t)$ and $\beta(t)$, we simplify
the rules as follows:

$(P):$ If $P(X|[x])\geq \alpha(t)$ ,
then  $x\in POS(X);$

$(B):$ If $\beta(t)<P(X|[x])< \alpha(t)$,  then
$x\in BND(X);$

$(N):$ If $P(X|[x])\leq \beta(t)$, then
$x\in NEG(X). $ $\Box$

In the following, we employ an example to illustrate that how to
compute $\alpha(t)$ and $\beta(t)$ by using a loss
function.

\begin{example}
Let $\lambda_{PP}(t)=0,  \lambda_{NN}(t)=0, \lambda_{BP}(t)\sim
U(2t+2, 4t+4), \lambda_{NP}(t)\sim U(3t+6, 5t+12),
\lambda_{PN}(t)\sim U(2t+14, 4t+20)$ and $ \lambda_{BN}(t)\sim
U(t+2, 3t+10).$ Then we have
\begin{eqnarray*}
\lambda_{PP}(t)&=&\lambda_{NN}(t)=0,\\
\lambda_{PN}(t)&=&\frac{a_{NN}(t)+b_{NN}(t)}{2}=\frac{2t+14+4t+20}{2}=3t+17,\\
\lambda_{BP}(t)&=&\frac{a_{BP}(t)+b_{BP}(t)}{2}=\frac{2t+2+4t+4}{2}=3t+3,\\
\lambda_{BN}(t)&=&\frac{a_{BN}(t)+b_{BN}(t)}{2}=\frac{t+2+3t+10}{2}=2t+6,\\
\lambda_{NP}(t)&=&\frac{a_{NP}(t)+b_{NP}(t)}{2}=\frac{3t+6+5t+12}{2}=4t+9.
\end{eqnarray*}

By Theorem 3.2, we have
\begin{eqnarray*}
\alpha(t)&=&\frac{\lambda_{PN}(t)-\lambda_{BN}(t)}{\lambda_{PN}(t)-\lambda_{BN}(t)+\lambda_{BP}(t)-\lambda_{PP}(t)}
=\frac{t+11}{4t+14};\\
\beta(t)&=&\frac{\lambda_{BN}(t)-\lambda_{NN}(t)}{\lambda_{BN}(t)-\lambda_{NN}(t)+\lambda_{NP}(t)-\lambda_{BP}(t)}
=\frac{2t+6}{3t+12}.
\end{eqnarray*}
\end{example}

\subsection{Normal processes-based DTRS}

In this subsection, we introduce the concept of normal processes for
DTRS.

\begin{definition}
Let $X(t)$ be a variation of the time $t$, the probability density
function of $X(t)$ is \makeatother $$f(x, \mu(t),
\sigma^{2}(t))=\frac{1}{\sigma(t)\sqrt{2\pi}}e^{-\frac{(x-\mu(t))^{2}}{2\sigma^{2}(t)}}.$$
Then $X(t)$ is called a normal process which is satisfied
mathematical expectations $\mu(t)$ and  variance $\sigma^{2}(t)$,
denoted as $X(t)\sim U(\mu(t), \sigma^{2}(t))$.
\end{definition}

In \cite{Liu9}, Liu et al. discussed DTRS when
loss functions are satisfied normal distributions, and there is a need
to study DTRS when loss functions are satisfied
normal processes.

In what follows, we discuss DTRS when loss
functions are satisfied normal processes. Suppose loss functions in
decision-theoretic rough set theory are satisfied normal processes as
follows:
\begin{eqnarray*}
\lambda_{PP}(t)&\sim& U(\mu_{PP}(t), \sigma_{PP}^{2}(t)),
\lambda_{BP}(t)\sim U(\mu_{BP}(t), \sigma_{BP}^{2}(t)),
\lambda_{NP}(t)\sim U(\mu_{NP}(t), \sigma_{NP}^{2}(t)),\\
\lambda_{NN}(t)&\sim& U(\mu_{NN}(t), \sigma_{NN}^{2}(t)),
\lambda_{BN}(t)\sim U(\mu_{BN}(t), \sigma_{BN}^{2}(t)),
\lambda_{PN}(t)\sim U(\mu_{PN}(t), \sigma_{PN}^{2}(t)).
\end{eqnarray*}

Suppose $\lambda(t)\sim U(\mu(t), \sigma^{2}(t))$, we have
$P(\mu(t)-\sigma(t)\leq\lambda(t)\leq \mu(t)+\sigma(t))\approx
0.6827, P(\mu(t)-2\sigma(t)\leq\lambda(t)\leq
\mu(t)+2\sigma(t))\approx 0.9545$ and
$P(\mu(t)-3\sigma(t)\leq\lambda(t)\leq \mu(t)+3\sigma(t))\approx
0.9973$. If we take three confidence intervals $[\mu(t)-n\sigma(t),
\mu(t)+n\sigma(t)](n=1,2,3)$ instead of $\lambda(t)$ for loss
function and suppose $\mu(t)-n\sigma(t)\geq 0$ for $\mu(t)$, then
the expected losses $R(a_{P}|[x])$, $R(a_{B}|[x])$ and
$R(a_{N}|[x])$ associated with taking the individual actions for an
object $x$ and $t\in T$ are shown as follows:
\begin{eqnarray*}
R(a_{P}|[x])=\lambda_{PP}(t)P(X|[x])+\lambda_{PN}(t)P(\neg
X|[x]);\\
R(a_{B}|[x])=\lambda_{BP}(t)P(X|[x])+\lambda_{BN}(t)P(\neg
X|[x]);\\
R(a_{N}|[x])=\lambda_{NP}(t)P(X|[x])+\lambda_{NN}(t)P(\neg X|[x]).
\end{eqnarray*}

The bayesian decision procedure suggests the local minimum-cost
decision rules:

$(P):$ If $R(a_{P}|[x])\leq R(a_{B}|[x])$ and $R(a_{P}|[x])\leq
R(a_{N}|[x])$, then $x\in POS(X);$

$(B):$ If $R(a_{B}|[x])\leq R(a_{P}|[x])$ and $R(a_{B}|[x])\leq
R(a_{N}|[x])$, then $x\in BND(X);$

$(N):$ If $R(a_{N}|[x])\leq R(a_{P}|[x])$ and $R(a_{N}|[x])\leq
R(a_{B}|[x])$, then $x\in NEG(X).$

Since $P(X|[x])+P(\neg X|[x])=1$, we take the values of $\alpha(t)$ and $
\beta(t)$ as follows:
\begin{eqnarray*}
\alpha(t)=\frac{\lambda_{PN}(t)-\lambda_{BN}(t)}{\lambda_{PN}(t)-\lambda_{BN}(t)+\lambda_{BP}(t)-\lambda_{PP}(t)},
\beta(t)=\frac{\lambda_{BN}(t)-\lambda_{NN}(t)}{\lambda_{BN}(t)-\lambda_{NN}(t)+\lambda_{NP}(t)-\lambda_{BP}(t)}.
\end{eqnarray*}
Concretely, on the basis of
$\alpha^{min}(t),\alpha^{max}(t),\beta^{min}(t)$ and $\beta^{max}(t)
$, we have the following results:
\begin{eqnarray*}
\alpha(t)\in[max\{\alpha^{min}(t),0\},min\{\alpha^{max}(t),1\}],
\beta(t)\in[max\{\beta^{min}(t),0\},min\{\beta^{max}(t),1\}],
\end{eqnarray*}where
\begin{eqnarray*}
\alpha^{min}(t)&=&\frac{[\mu_{PN}(t)-n\sigma_{PN}(t)]-[\mu_{BN}(t)+n\sigma_{BN}(t)]}{[\mu_{PN}(t)+n\sigma_{PN}(t)]-[\mu_{BN}(t)-n\sigma_{BN}(t)]+[\mu_{BP}(t)+n\sigma_{BP}(t)]-[\mu_{PP}(t)-n\sigma_{PP}(t)]},\\
\alpha^{max}(t)&=&\frac{[\mu_{PN}(t)+n\sigma_{PN}(t)]-[\mu_{BN}(t)-n\sigma_{BN}(t)]}{[\mu_{PN}(t)-n\sigma_{PN}(t)]-[\mu_{BN}(t)+n\sigma_{BN}(t)]+[\mu_{BP}(t)-n\sigma_{BP}(t)]-[\mu_{PP}(t)+n\sigma_{PP}(t)]},\\
\beta^{min}(t)&=&\frac{[\mu_{BN}(t)-n\sigma_{BN}(t)]-[\mu_{NN}(t)+n\sigma_{NN}(t)]}{[\mu_{BN}(t)+n\sigma_{BN}(t)]-[\mu_{NN}(t)-n\sigma_{NN}(t)]+[\mu_{NP}(t)+n\sigma_{NP}(t)]-[\mu_{BP}(t)-n\sigma_{BP}(t)]},
\\
\beta^{max}(t)&=&\frac{[\mu_{BN}(t)+n\sigma_{BN}(t)]-[\mu_{NN}(t)-n\sigma_{NN}(t)]}{[\mu_{BN}(t)-n\sigma_{BN}(t)]-[\mu_{NN}(t)+n\sigma_{NN}(t)]+[\mu_{NP}(t)-n\sigma_{NP}(t)]-[\mu_{BP}(t)+n\sigma_{BP}(t)]}.
\end{eqnarray*}

By using the thresholds $\alpha(t)$ and $\beta(t)$, we
simplify the rules as follows:

$(P):$ If $P(X|[x])\geq \alpha(t)$,
then  $x\in POS(X);$

$(B):$ If $\beta(t)<P(X|[x])< \alpha(t)$,  then
$x\in BND(X);$

$(N):$ If $P(X|[x])\leq \beta(t)$, then
$x\in NEG(X).$

\begin{example}
Let $\lambda_{PP}(t)\sim U(\frac{3t+2}{2},
\frac{t^{2}+4n+4}{4n^{2}}), \lambda_{BP}(t)\sim U(\frac{5t+8}{2},
\frac{t^{2}+4n+4}{4n^{2}}), \lambda_{NP}(t)\sim U(\frac{7t+14}{2},
\frac{t^{2}+4n+4}{4n^{2}}), \lambda_{NN}(t)\sim U(\frac{3t+2}{2},
\frac{t^{2}+4n+4}{4n^{2}}), \lambda_{BN}(t)\sim U(\frac{5t+8}{2},
\frac{t^{2}+4n+4}{4n^{2}})$ and $ \lambda_{PN}(t)\sim
U(\frac{7t+18}{2}, \frac{t^{2}+4n+4}{4n^{2}}).$ Then we have
\begin{eqnarray*}
\alpha^{min}(t)=\frac{3}{4t+12},
\beta^{min}(t)=\frac{1}{4t+10},
\alpha^{max}(t)=\frac{2t+7}{4},\beta^{max}(t)=\frac{2t+5}{1}.
\end{eqnarray*}
\end{example}

Consequently, we compute the thresholds $\alpha(t)$ and $\beta(t)$ when loss functions are two special cases as follows.

$(1)$ Considering $\lambda^{1}_{PP}(t)=\mu_{PP}(t)-n\sigma_{PP}(t),$
$\lambda^{1}_{BP}(t)=\mu_{BP}(t)-n\sigma_{BP}(t),$
$\lambda^{1}_{NP}(t)=\mu_{NP}(t)-n\sigma_{NP}(t),$
$\lambda^{1}_{NN}(t)=\mu_{NN}(t)-n\sigma_{NN}(t),$
$\lambda^{1}_{BN}(t)=\mu_{BN}(t)-n\sigma_{BN}(t)$ and
$\lambda^{1}_{PN}(t)=\mu_{PN}(t)-n\sigma_{PN}(t),$ then we have
\begin{eqnarray*}
\alpha^{1}(t)&=&\frac{[\mu_{PN}(t)-n\sigma_{PN}(t)]-[\mu_{BN}(t)-n\sigma_{BN}(t)]}{[\mu_{PN}(t)-n\sigma_{PN}(t)]-[\mu_{BN}(t)-n\sigma_{BN}(t)]+[\mu_{BP}(t)-n\sigma_{BP}(t)]-[\mu_{PP}(t)-n\sigma_{PP}(t)]};\\
\beta^{1}(t)&=&\frac{[\mu_{BN}(t)-n\sigma_{BN}(t)]-[\mu_{NN}(t)-n\sigma_{NN}(t)]}{[\mu_{BN}(t)-n\sigma_{BN}(t)]-[\mu_{NN}(t)-n\sigma_{NN}(t)]+[\mu_{NP}(t)-n\sigma_{NP}(t)]-[\mu_{BP}(t)-n\sigma_{BP}(t)]}.
\end{eqnarray*}

$(2)$ Considering $\lambda^{2}_{PP}(t)=\mu_{PP}(t)+n\sigma_{PP}(t),$
$\lambda^{2}_{BP}(t)=\mu_{BP}(t)+n\sigma_{BP}(t),$
$\lambda^{2}_{NP}(t)=\mu_{NP}(t)+n\sigma_{NP}(t),$
$\lambda^{2}_{NN}(t)=\mu_{NN}(t)+n\sigma_{NN}(t),$
$\lambda^{2}_{BN}(t)=\mu_{BN}(t)+n\sigma_{BN}(t),$
$\lambda^{2}_{PN}(t)=\mu_{PN}(t)+n\sigma_{PN}(t),$ then we have
\begin{eqnarray*}
\alpha^{2}(t)&=&\frac{[\mu_{PN}(t)+n\sigma_{PN}(t)]-[\mu_{BN}(t)+n\sigma_{BN}(t)]}{[\mu_{PN}(t)+n\sigma_{PN}(t)]-[\mu_{BN}(t)+n\sigma_{BN}(t)]+[\mu_{BP}(t)+n\sigma_{BP}(t)]-[\mu_{PP}(t)+n\sigma_{PP}(t)]};\\
\beta^{2}(t)&=&\frac{[\mu_{BN}(t)+n\sigma_{BN}(t)]-[\mu_{NN}(t)+n\sigma_{NN}(t)]}{[\mu_{BN}(t)+n\sigma_{BN}(t)]-[\mu_{NN}(t)+n\sigma_{NN}(t)]+[\mu_{NP}(t)+n\sigma_{NP}(t)]-[\mu_{BP}(t)+n\sigma_{BP}(t)]}.
\end{eqnarray*}

On the basis of the above results, we have that
$\alpha^{1}(t),\alpha^{2}(t)\in
[max\{\alpha^{min}(t),0\},min\{\alpha^{max}(t),1\}]$ and
$\beta^{1}(t),\beta^{2}(t)\in
[max\{\beta^{min}(t),0\},min\{\beta^{max}(t),1\}].$

\section{Time-dependent interval sets-based DTRS}

In this section, we discuss DTRS when loss
functions are time-dependent interval sets. Firstly, we present the
concept of time-dependent interval sets for DTRS.

\begin{definition}
Let $\lambda(t)=[a(t),b(t)]$, where $a(t)$ and $b(t)$ are variations
of the time $t$, then $\lambda(t)$ is called a time-dependent interval set.
\end{definition}

In what follows, we employ Table 3 to illustrate loss functions which
are time-dependent interval sets.

\begin{table}[htbp]\renewcommand{\arraystretch}{1.5}
\caption{Time-dependent loss function.}
 \tabcolsep0.5in
\begin{tabular}{c c c }
\hline $Action$  & $X(P)$ &$\neg X(N)$\\
\hline $a_{P}$ &
$\lambda_{PP}(t)=[\lambda^{min}_{PP}(t),\lambda^{max}_{PP}(t)]$&
$\lambda_{PN}(t)=[\lambda^{min}_{PN}(t),\lambda^{max}_{PN}(t)]$  \\
$a_{B}$ &
$\lambda_{BP}(t)=[\lambda^{min}_{BP}(t),\lambda^{max}_{BP}(t)]$&
$\lambda_{BN}(t)=[\lambda^{min}_{BN}(t),\lambda^{max}_{BN}(t)]$ \\
$a_{N}$ &
$\lambda_{NP}(t)=[\lambda^{min}_{NP}(t),\lambda^{max}_{NP}(t)]$&
$\lambda_{NN}(t)=[\lambda^{min}_{NN}(t),\lambda^{max}_{NN}(t)]$  \\
\hline
\end{tabular}
\end{table}

In Table 3,
$\lambda_{PP}(t),\lambda_{BP}(t),\lambda_{NP}(t),\lambda_{PN}(t),\lambda_{BN}(t)$
and $\lambda_{NN}(t)$ are time-dependent interval sets. Below, we discuss
DTRS when loss functions are the
lower and upper bounds of time-dependent interval sets.

On one hand,
$\lambda^{min}_{PP}(t),\lambda^{min}_{BP}(t),\lambda^{min}_{NP}(t),\lambda^{min}_{PN}(t),
\lambda^{min}_{BN}(t)$ and $\lambda^{min}_{NN}(t)$ denote the lower
bounds of time-dependent interval sets in Table 3. The expected losses
$R^{opt}(a_{P}|[x])$, $R^{opt}(a_{B}|[x])$ and $R^{opt}(a_{N}|[x])$
associated with taking the individual actions for an object $x$ and
$t\in T$ are shown as follows:
\begin{eqnarray*}
R^{opt}(a_{P}|[x])&=&\lambda^{min}_{PP}(t)P(X|[x])+\lambda^{min}_{PN}(t)P(\neg
X|[x]);\\
R^{opt}(a_{B}|[x])&=&\lambda^{min}_{BP}(t)P(X|[x])+\lambda^{min}_{BN}(t)P(\neg
X|[x]);\\
R^{opt}(a_{N}|[x])&=&\lambda^{min}_{NP}(t)P(X|[x])+\lambda^{min}_{NN}(t)P(\neg
X|[x]).
\end{eqnarray*}

The bayesian decision procedure suggests the local minimum-cost
decision rules:

$(P):$ If $R^{opt}(a_{P}|[x])\leq R^{opt}(a_{B}|[x])$ and
$R^{opt}(a_{P}|[x])\leq R^{opt}(a_{N}|[x]),$ then $x\in POS(X);$

$(B):$ If $R^{opt}(a_{B}|[x])\leq R^{opt}(a_{P}|[x])$ and
$R^{opt}(a_{B}|[x])\leq R^{opt}(a_{N}|[x])$, then $x\in BND(X);$

$(N):$ If $R^{opt}(a_{N}|[x])\leq R^{opt}(a_{P}|[x])$ and
$R^{opt}(a_{N}|[x])\leq R^{opt}(a_{B}|[x])$, then $x\in NEG(X).$

Suppose $0\leq
\lambda^{min}_{PP}(t)\leq\lambda^{min}_{BP}(t)\leq\lambda^{min}_{NP}(t)$
and $0\leq \lambda^{min}_{NN}(t)\leq
\lambda^{min}_{BN}(t)\leq\lambda^{min}_{PN}(t)$ for $t\in T$. Since
$P(X|[x])+P(\neg X|[x])=1$, we simplify the rules as follows:

$(P):$ If $P(X|[x])\geq \alpha^{opt}(t)$, then  $x\in POS(X);$

$(B):$ If $\beta^{opt}(t)<P(X|[x])< \alpha^{opt}(t)$, then $x\in BND(X);$

$(N):$ If $P(X|[x])\leq \beta^{opt}(t)$, then $x\in NEG(X),$
\noindent where
\begin{eqnarray*}
\alpha^{opt}(t)=\frac{\lambda^{min}_{PN}(t)-\lambda^{min}_{BN}(t)}{\lambda^{min}_{PN}(t)-\lambda^{min}_{BN}(t)+\lambda^{min}_{BP}(t)-\lambda^{min}_{PP}(t)},
\beta^{opt}(t)=\frac{\lambda^{min}_{BN}(t)-\lambda^{min}_{NN}(t)}{\lambda^{min}_{BN}(t)-\lambda^{min}_{NN}(t)+\lambda^{min}_{NP}(t)-\lambda^{min}_{BP}(t)},
\end{eqnarray*}
\begin{example} Let
$\lambda_{PP}(t)=[t,2t+2], \lambda_{PN}(t)=[4t+8,4t+10],
\lambda_{BP}(t)=[2t+3,2t+5], \lambda_{BN}(t)=[3t+2,3t+6],
\lambda_{NP}(t)=[3t+6,3t+8]$ and $ \lambda_{NN}(t)=[2t,2t+2]$. Then
we have
\begin{eqnarray*}
\alpha^{opt}(t)&=&\frac{\lambda^{min}_{PN}(t)-\lambda^{min}_{BN}(t)}{\lambda^{min}_{PN}(t)-\lambda^{min}_{BN}(t)+\lambda^{min}_{BP}(t)-\lambda^{min}_{PP}(t)}=\frac{t+6}{2t+9};\\
\beta^{opt}(t)&=&\frac{\lambda^{min}_{BN}(t)-\lambda^{min}_{NN}(t)}{\lambda^{min}_{BN}(t)-\lambda^{min}_{NN}(t)+\lambda^{min}_{NP}(t)-\lambda^{min}_{BP}(t)}=\frac{t+2}{2t+5}.
\end{eqnarray*}
\end{example}

On the other hand,
$\lambda^{max}_{PP}(t),\lambda^{max}_{BP}(t),\lambda^{max}_{NP}(t),\lambda^{max}_{PN}(t),
\lambda^{max}_{BN}(t)$ and $\lambda^{max}_{NN}(t)$ are upper bounds
of time-dependent interval sets in Table 3. We have the expected losses
$R^{pes}(a_{P}|[x])$, $R^{pes}(a_{B}|[x])$ and $R^{pes}(a_{N}|[x])$
associated with taking the individual actions for an object $x$ and
$t\in T$ as follows:
\begin{eqnarray*}
R^{pes}(a_{P}|[x])&=&\lambda^{max}_{PP}(t)P(X|[x])+\lambda^{max}_{PN}(t)P(\neg
X|[x]);\\
R^{pes}(a_{B}|[x])&=&\lambda^{max}_{BP}(t)P(X|[x])+\lambda^{max}_{BN}(t)P(\neg
X|[x]);\\
R^{pes}(a_{N}|[x])&=&\lambda^{max}_{NP}(t)P(X|[x])+\lambda^{max}_{NN}(t)P(\neg
X|[x]).
\end{eqnarray*}

The bayesian decision procedure suggests the local minimum-cost
decision rules:

$(P'):$ If $R^{pes}(a_{P}|[x])\leq R^{pes}(a_{B}|[x])$ and
$R^{pes}(a_{P}|[x])\leq R^{pes}(a_{N}|[x]),$ then $x\in POS(X);$

$(B'):$ If $R^{pes}(a_{B}|[x])\leq R^{pes}(a_{P}|[x])$ and
$R^{pes}(a_{B}|[x])\leq R^{pes}(a_{N}|[x]),$ then $x\in BND(X);$

$(N'):$ If $R^{pes}(a_{N}|[x])\leq R^{pes}(a_{P}|[x])$ and
$R^{pes}(a_{N}|[x])\leq R^{pes}(a_{B}|[x]),$ then $x\in NEG(X).$

Suppose $0\leq
\lambda^{max}_{PP}(t)\leq\lambda^{max}_{BP}(t)\leq\lambda^{max}_{NP}(t)$
and $0\leq \lambda^{max}_{NN}(t)\leq
\lambda^{max}_{BN}(t)\leq\lambda^{max}_{PN}(t)$ for $t\in T$. Since
$P(X|[x])+P(\neg X|[x])=1$, we simplify the rules as follows:

$(P'):$ If $P(X|[x])\geq \alpha^{pes}(t)$, then  $x\in POS(X);$

$(B'):$ If $\beta^{pes}(t)<P(X|[x])< \alpha^{pes}(t)$, then $x\in BND(X);$

$(N'):$ If $P(X|[x])\leq \beta^{pes}(t)$, then $x\in NEG(X),$
\noindent where
\begin{eqnarray*}
\alpha^{pes}(t)=\frac{\lambda^{max}_{PN}(t)-\lambda^{max}_{BN}(t)}{\lambda^{max}_{PN}(t)-\lambda^{max}_{BN}(t)+\lambda^{max}_{BP}(t)-\lambda^{max}_{PP}(t)},
\beta^{pes}(t)=\frac{\lambda^{max}_{BN}(t)-\lambda^{max}_{NN}(t)}{\lambda^{max}_{BN}(t)-\lambda^{max}_{NN}(t)+\lambda^{max}_{NP}(t)-\lambda^{max}_{BP}(t)}.
\end{eqnarray*}
\begin{example} Let
$\lambda_{PP}(t)=[t,2t+2], \lambda_{PN}(t)=[4t+8,4t+10],
\lambda_{BP}(t)=[2t+3,2t+5], \lambda_{BN}(t)=[3t+2,3t+6],
\lambda_{NP}(t)=[3t+6,3t+8]$ and $ \lambda_{NN}(t)=[2t,2t+2]$. Then
we have
\begin{eqnarray*}
\alpha^{pes}(t)&=&\frac{\lambda^{max}_{PN}(t)-\lambda^{max}_{BN}(t)}{\lambda^{max}_{PN}(t)-\lambda^{max}_{BN}(t)+\lambda^{max}_{BP}(t)-\lambda^{max}_{PP}(t)}=\frac{t+4}{t+7};\\
\beta^{pes}(t)&=&\frac{\lambda^{max}_{BN}(t)-\lambda^{max}_{NN}(t)}{\lambda^{max}_{BN}(t)-\lambda^{max}_{NN}(t)+\lambda^{max}_{NP}(t)-\lambda^{max}_{BP}(t)}=\frac{t+4}{2t+7}.
\end{eqnarray*}
\end{example}

In general, by taking $\lambda^{\ast}_{PP}(t)\in
\lambda_{PP}(t),\lambda^{\ast}_{BP}(t)\in
\lambda_{BP}(t),\lambda^{\ast}_{NP}(t)\in
\lambda_{NP}(t),\lambda^{\ast}_{PN}(t)\in \lambda_{PN}(t),
\lambda^{\ast}_{BN}(t)\in \lambda_{BN}(t)$ and
$\lambda^{\ast}_{NN}(t)\in \lambda_{NN}(t)$, we have the expected
losses $R(a_{P}|[x])$, $R(a_{B}|[x])$ and $R(a_{N}|[x])$ associated
with taking the individual actions for an object $x$ and $t\in T$ as
follows:
\begin{eqnarray*}
R(a_{P}|[x])&=&\lambda^{\ast}_{PP}(t)P(X|[x])+\lambda^{\ast}_{PN}(t)P(\neg
X|[x]);\\
R(a_{B}|[x])&=&\lambda^{\ast}_{BP}(t)P(X|[x])+\lambda^{\ast}_{BN}(t)P(\neg
X|[x]);\\
R(a_{N}|[x])&=&\lambda^{\ast}_{NP}(t)P(X|[x])+\lambda^{\ast}_{NN}(t)P(\neg
X|[x]).
\end{eqnarray*}

The bayesian decision procedure suggests the local minimum-cost
decision rules:

$(P''):$ If $R(a_{P}|[x])\leq R(a_{B}|[x])$ and $R(a_{P}|[x])\leq
R(a_{N}|[x]),$ then $x\in POS(X);$

$(B''):$ If $R(a_{B}|[x])\leq R(a_{P}|[x])$ and $R(a_{B}|[x])\leq
R(a_{N}|[x]),$ then $x\in BND(X);$

$(N''):$ If $R(a_{N}|[x])\leq R(a_{P}|[x])$ and $R(a_{N}|[x])\leq
R(a_{B}|[x]),$ then $x\in NEG(X).$

Suppose $0\leq
\lambda^{\ast}_{PP}(t)\leq\lambda^{\ast}_{BP}(t)\leq\lambda^{\ast}_{NP}(t)$
and $0\leq \lambda^{\ast}_{NN}(t)\leq
\lambda^{\ast}_{BN}(t)\leq\lambda^{\ast}_{PN}(t)$ for $t\in T$.
Since $P(X|[x])+P(\neg X|[x])=1$, we simplify the rules as follows:

$(P''):$ If $P(X|[x])\geq \alpha(t)$,
then  $x\in POS(X);$

$(B''):$ If $\beta(t)<P(X|[x])< \alpha(t)$, then
$x\in BND(X);$

$(N''):$ If $P(X|[x])\leq \beta(t)$,
then $x\in NEG(X),$
\noindent where
\begin{eqnarray*}
\alpha(t)=\frac{\lambda^{\ast}_{PN}(t)-\lambda^{\ast}_{BN}(t)}{\lambda^{\ast}_{PN}(t)-\lambda^{\ast}_{BN}(t)+\lambda^{\ast}_{BP}(t)-\lambda^{\ast}_{PP}(t)},
\beta(t)=\frac{\lambda^{\ast}_{BN}(t)-\lambda^{\ast}_{NN}(t)}{\lambda^{\ast}_{BN}(t)-\lambda^{\ast}_{NN}(t)+\lambda^{\ast}_{NP}(t)-\lambda^{\ast}_{BP}(t)}.
\end{eqnarray*}
\begin{theorem}
Let $0\leq
\lambda^{min}_{PP}(t)\leq\lambda^{max}_{PP}(t)\leq\lambda^{min}_{BP}(t)\leq
\lambda^{max}_{BP}(t)\leq\lambda^{min}_{NP}(t)\leq\lambda^{max}_{NP}(t)$
and $0\leq
\lambda^{min}_{NN}(t)\leq\lambda^{max}_{NN}(t)\leq\lambda^{min}_{BN}(t)\leq
\lambda^{max}_{BN}(t)\leq\lambda^{min}_{PN}(t)\leq\lambda^{max}_{PN}(t)$,
where $t\in T$. Then
\begin{eqnarray*}
(1)&&
\alpha(t)\in[\frac{\lambda^{min}_{PN}(t)-\lambda^{max}_{BN}(t)}{\lambda
^{max}_{PN}(t)-\lambda^{min}_{BN}(t)+\lambda^{max}_{BP}(t)-\lambda^{min}_{PP}(t)},
min\{\frac{\lambda^{max}_{PN}(t)-\lambda^{min}_{BN}(t)}{\lambda^{min}_{PN}(t)-\lambda^{max}_{BN}(t)
+\lambda^{min}_{BP}(t)-\lambda^{max}_{PP}(t)},1\}];
\\
(2)&&
\beta(t)\in[\frac{\lambda^{min}_{BN}(t)-\lambda^{max}_{NN}(t)}{\lambda
^{max}_{BN}(t)-\lambda^{min}_{NN}(t)+\lambda^{max}_{NP}(t)-\lambda^{min}_{BP}(t)},
min\{\frac{\lambda^{max}_{BN}(t)-\lambda^{min}_{NN}(t)}{\lambda^{min}_{BN}(t)-\lambda^{max}_{NN}(t)
+\lambda^{min}_{NP}(t)-\lambda^{max}_{BP}(t)},1\}].
\end{eqnarray*}
\end{theorem}

\noindent\textbf{Proof.} $(1)$ Since $0\leq
\lambda^{min}_{PN}(t)\leq\lambda^{max}_{PP}(t)\leq\lambda^{min}_{BP}(t)\leq
\lambda^{max}_{BP}(t)\leq\lambda^{min}_{NP}(t)\leq\lambda^{max}_{NP}(t)$
and $0\leq
\lambda^{min}_{NN}(t)\leq\lambda^{max}_{NN}(t)\leq\lambda^{min}_{BN}(t)\leq
\lambda^{max}_{BN}(t)\leq\lambda^{min}_{PN}(t)\leq\lambda^{max}_{PN}(t)$,
we have
\begin{eqnarray*}
&&\lambda^{\ast}_{PN}(t)-\lambda^{\ast}_{BN}(t)\geq 0,
\lambda^{\ast}_{BP}(t)-\lambda^{\ast}_{PP}(t)\geq 0,\\
&&\lambda^{min}_{PN}(t)-\lambda^{max}_{PN}(t)\leq
\lambda^{\ast}_{PN}(t)-\lambda^{\ast}_{BN}(t)\leq
\lambda^{max}_{PN}(t)-\lambda^{min}_{PN}(t),\\
&&\lambda^{min}_{BP}(t)-\lambda^{max}_{PP}(t)\leq
\lambda^{\ast}_{BP}(t)-\lambda^{\ast}_{PP}(t)\leq
\lambda^{max}_{BP}(t)-\lambda^{min}_{PP}(t).
\end{eqnarray*}
It implies that
\begin{eqnarray*}
\lambda^{min}_{PN}(t)-\lambda^{max}_{BN}(t)+
\lambda^{min}_{BP}(t)-\lambda^{max}_{PP}(t)&\leq&
\lambda^{\ast}_{PN}(t)-\lambda^{\ast}_{BN}(t)+
\lambda^{\ast}_{BP}(t)-\lambda^{\ast}_{PP}(t)\\&\leq&
\lambda^{max}_{PN}(t)-\lambda^{min}_{BN}(t)+
\lambda^{max}_{BP}(t)-\lambda^{min}_{PP}(t).
\end{eqnarray*}
It follows that
\begin{eqnarray*}
\frac{\lambda^{min}_{PN}(t)-\lambda^{max}_{BN}(t)}{\lambda
^{max}_{PN}(t)-\lambda^{min}_{BN}(t)+\lambda^{max}_{BP}(t)-\lambda^{min}_{PP}(t)}&\leq&
\frac{\lambda^{\ast}_{PN}(t)-\lambda^{\ast}_{BN}(t)}{\lambda^{\ast}
_{PN}(t)-\lambda^{\ast}_{BN}(t)+\lambda^{\ast}_{BP}(t)-\lambda^{\ast}_{PP}(t)}\\&\leq&
\frac{\lambda^{max}_{PN}(t)-\lambda^{min}_{BN}(t)}{\lambda
^{min}_{PN}(t)-\lambda^{max}_{BN}(t)+\lambda^{min}_{BP}(t)-\lambda^{max}_{PP}(t)}.
\end{eqnarray*}
Obviously, we have
$$\alpha(t),
\frac{\lambda^{min}_{PN}(t)-\lambda^{max}_{BN}(t)}{\lambda
^{max}_{PN}(t)-\lambda^{min}_{BN}(t)+\lambda^{max}_{BP}(t)-\lambda^{min}_{PP}(t)},
\frac{\lambda^{max}_{PN}(t)-\lambda^{min}_{BN}(t)}{\lambda
^{min}_{PN}(t)-\lambda^{max}_{BN}(t)+\lambda^{min}_{BP}(t)-\lambda^{max}_{PP}(t)}\in
[0,1].$$ Therefore,
$$\alpha(t)\in[\frac{\lambda^{min}_{PN}(t)-\lambda^{max}_{BN}(t)}{\lambda
^{max}_{PN}(t)-\lambda^{min}_{BN}(t)+\lambda^{max}_{BP}(t)-\lambda^{min}_{PP}(t)},
min\{\frac{\lambda^{max}_{PN}(t)-\lambda^{min}_{BN}(t)}{\lambda^{min}_{PN}(t)-\lambda^{max}_{BN}(t)
+\lambda^{min}_{BP}(t)-\lambda^{max}_{PP}(t)},1\}].$$

$(2)$ Since $0\leq
\lambda^{min}_{PP}(t)\leq\lambda^{max}_{PP}(t)\leq\lambda^{min}_{BP}(t)\leq
\lambda^{max}_{BP}(t)\leq\lambda^{min}_{NP}(t)\leq\lambda^{max}_{NP}(t)$
and $0\leq
\lambda^{min}_{NN}(t)\leq\lambda^{max}_{NN}(t)\leq\lambda^{min}_{BN}(t)\leq
\lambda^{max}_{BN}(t)\leq\lambda^{min}_{PN}(t)\leq\lambda^{max}_{PN}(t)$,
we have
\begin{eqnarray*}
&&\lambda^{\ast}_{NP}(t)-\lambda^{\ast}_{BP}(t)\geq 0,
\lambda^{\ast}_{BN}(t)-\lambda^{\ast}_{NN}(t)\geq 0,\\
&&\lambda^{min}_{NP}(t)-\lambda^{max}_{BP}(t)\leq
\lambda^{\ast}_{NP}(t)-\lambda^{\ast}_{BP}(t)\leq
\lambda^{max}_{NP}(t)-\lambda^{min}_{BP}(t),\\
&&\lambda^{min}_{BN}(t)-\lambda^{max}_{NN}(t)\leq
\lambda^{\ast}_{BN}(t)-\lambda^{\ast}_{NN}(t)\leq
\lambda^{max}_{BN}(t)-\lambda^{min}_{NN}(t).
\end{eqnarray*}
It implies that
\begin{eqnarray*}
\lambda^{min}_{NP}(t)-\lambda^{max}_{BP}(t)+
\lambda^{min}_{BN}(t)-\lambda^{max}_{NN}(t)&\leq&
\lambda^{\ast}_{NP}(t)-\lambda^{\ast}_{NN}(t)+
\lambda^{\ast}_{BN}(t)-\lambda^{\ast}_{NN}(t)\\&\leq&
\lambda^{max}_{NP}(t)-\lambda^{min}_{BP}(t)+
\lambda^{max}_{BN}(t)-\lambda^{min}_{NN}(t).
\end{eqnarray*}
It follows that
\begin{eqnarray*}
\frac{\lambda^{min}_{BN}(t)-\lambda^{max}_{NN}(t)}{\lambda
^{max}_{BN}(t)-\lambda^{min}_{NN}(t)+\lambda^{max}_{NP}(t)-\lambda^{min}_{BP}(t)}&\leq&
\frac{\lambda^{\ast}_{BN}(t)-\lambda^{\ast}_{NN}(t)}{\lambda^{\ast}
_{BN}(t)-\lambda^{\ast}_{NN}(t)+\lambda^{\ast}_{NP}(t)-\lambda^{\ast}_{BP}(t)}\\&\leq&
\frac{\lambda^{max}_{BN}(t)-\lambda^{min}_{NN}(t)}{\lambda
^{min}_{BN}(t)-\lambda^{max}_{NN}(t)+\lambda^{min}_{NP}(t)-\lambda^{max}_{BP}(t)}.
\end{eqnarray*} Obviously, we have
$$\beta(t),
\frac{\lambda^{min}_{BN}(t)-\lambda^{max}_{NN}(t)}{\lambda
^{max}_{BN}(t)-\lambda^{min}_{NN}(t)+\lambda^{max}_{NP}(t)-\lambda^{min}_{BP}(t)},
\frac{\lambda^{max}_{BN}(t)-\lambda^{min}_{NN}(t)}{\lambda
^{min}_{BN}(t)-\lambda^{max}_{NN}(t)+\lambda^{min}_{NP}(t)-\lambda^{max}_{BP}(t)}\in
[0,1].$$ Therefore,
$$\beta(t)\in[\frac{\lambda^{min}_{BN}(t)-\lambda^{max}_{NN}(t)}{\lambda
^{max}_{BN}(t)-\lambda^{min}_{NN}(t)+\lambda^{max}_{NP}(t)-\lambda^{min}_{BP}(t)},
min\{\frac{\lambda^{max}_{BN}(t)-\lambda^{min}_{NN}(t)}{\lambda^{min}_{BN}(t)
-\lambda^{max}_{NN}(t)
+\lambda^{min}_{NP}(t)-\lambda^{max}_{BP}(t)},1\}].\Box$$

\section{Time-dependent fuzzy numbers-based DTRS}

In this section, we investigate DTRS when loss
functions are time-dependent fuzzy numbers. We introduce the
concepts of time-dependent fuzzy numbers and cut sets for DTRS.

\begin{definition}
Let $\mu_{\widetilde{A}(t)}$ be a mapping from $U$ to $[0,1]$ such
as $\mu_{\widetilde{A}(t)}: U\longrightarrow [0,1]:$ $
x\longrightarrow \mu_{\widetilde{A}(t)}, $ where $t\in T$,
$\mu_{\widetilde{A}(t)}$ is the membership function of
$\widetilde{A}(t)$, $\mu_{\widetilde{A}(t)}(x)$ is the membership
degree of $x$ to $\widetilde{A}(t)$, denoted as
$\widetilde{A}(t)=\{(x,\mu_{\widetilde{A}(t)}(x))|x\in U\},$ then
$\widetilde{A}(t)$ is called a time-dependent fuzzy number.
\end{definition}

\begin{example}
Let $\widetilde{A}(t)$ be a time-dependent fuzzy number, where
$\widetilde{A}(t)=\frac{t+1}{1-\frac{1}{t^{2}}}+\frac{2t+1}{1-\frac{1}{2t}}
+\frac{t+3}{1-\frac{1}{t+1}}+\frac{4t+1}{1-\frac{1}{3t+1}}+
\frac{2t-1}{1-\frac{2}{2t+1}}+\frac{4t-1}{1-\frac{3}{3t+1}}
+\frac{4t^{2}+1}{1-\frac{1}{3t+1}}+
\frac{2t^{2}+1}{1-\frac{2}{2t+1}}+\frac{4t^{2}+2t-1}{1-\frac{3}{3t^{2}+1}}.$
By Definition 5.1, we have that
$\mu_{\widetilde{A}(t)}(t+1)=1-\frac{1}{t^{2}}$ and $
\mu_{\widetilde{A}(t)}(2t^{2}+1)=1-\frac{2}{2t+1}.$
\end{example}

\begin{definition}
Let $\widetilde{A}(t)\in F(X),\forall\eta(t)\in[0,1]$, where $t\in
T$, then

$(1)$ $\widetilde{A}_{\eta}(t)=\{x|x\in
U,\mu_{\widetilde{A}(t)}\geq \eta(t)\}$ is referred to as a
$\eta(t)$-cut set of $\widetilde{A}(t)$;

$(2)$ $\widetilde{A}_{\eta^{>}}(t)=\{x|x\in U,\mu_{\widetilde{A}(t)}>
\eta(t)\}$ is referred to as a strong $\eta(t)$-cut set of
$\widetilde{A}(t)$.
\end{definition}

\begin{example}
Let $\widetilde{A}(t)$ be a time-dependent fuzzy number, where
$\widetilde{A}(t)=\frac{t+1}{1-\frac{1}{t}}+\frac{2t+1}{1-\frac{1}{2t}}
+\frac{t+3}{1-\frac{1}{t}}+\frac{4t+1}{1-\frac{1}{2t}}+
\frac{2t-1}{1-\frac{1}{2t+1}}+\frac{4t-1}{1-\frac{1}{2t}}
+\frac{4t^{2}+1}{1-\frac{1}{t}}+
\frac{2t^{2}+1}{1-\frac{1}{2t+1}}+\frac{4t^{2}+2t-1}{1-\frac{1}{t}}.$
By Definition 5.3, we have that
$\widetilde{A}_{1-\frac{1}{2t}}(t)=\{2t+1,4t+1, 4t-1,
2t-1,2t^{2}+1\}$ and
$\widetilde{A}_{(1-\frac{1}{2t})^{>}}(t)=\{2t-1,2t^{2}+1\}.$
\end{example}

In what follows, we employ Table 4 to illustrate loss functions which
are time-dependent fuzzy numbers.

\begin{table}[htbp]\renewcommand{\arraystretch}{1.5}
\caption{Time-dependent fuzzy loss function.}
 \tabcolsep0.5in
\begin{tabular}{c c c }
\hline $Action$  & $X(P)$ &$\neg X(N)$\\
\hline $a_{P}$ &
$\widetilde{\lambda^{\eta}}_{PP}(t)=[PP^{L}_{\eta}(t),PP^{U}_{\eta}(t)]$&
$\widetilde{\lambda^{\eta}}_{PN}(t)=[PN^{L}_{\eta}(t),PN^{U}_{\eta}(t)]$  \\
$a_{B}$ &
$\widetilde{\lambda^{\eta}}_{BP}(t)=[BP^{L}_{\eta}(t),BP^{U}_{\eta}(t)]$&
$\widetilde{\lambda^{\eta}}_{BN}(t)=[BN^{L}_{\eta}(t),BN^{U}_{\eta}(t)]$ \\
$a_{N}$ &
$\widetilde{\lambda^{\eta}}_{NP}(t)=[NP^{L}_{\eta}(t),NP^{U}_{\eta}(t)]$&
$\widetilde{\lambda^{\eta}}_{NN}(t)=[NN^{L}_{\eta}(t),NN^{U}_{\eta}(t)]$  \\
\hline
\end{tabular}
\end{table}

In Table 4, $\widetilde{\lambda^{\eta}}_{PP}(t),
\widetilde{\lambda^{\eta}}_{BP}(t),
\widetilde{\lambda^{\eta}}_{NP}(t),
\widetilde{\lambda^{\eta}}_{PN}(t),
\widetilde{\lambda^{\eta}}_{BN}(t)$ and $
\widetilde{\lambda^{\eta}}_{NN}(t)$ are time-dependent fuzzy numbers.
Below, we discuss DTRS when loss functions are
the lower and upper bounds of time-dependent fuzzy numbers.

On one hand,
$PP^{L}_{\eta}(t),BP^{L}_{\eta}(t),NP^{L}_{\eta}(t),PN^{L}_{\eta}(t)
,BN^{L}_{\eta}(t)$ and $NN^{L}_{\eta}(t)$ denote the lower bounds of
time-dependent fuzzy numbers in Table 4. We have the expected losses
$R^{opt}(a_{P}|[x])$, $R^{opt}(a_{B}|[x])$ and $R^{opt}(a_{N}|[x])$
associated with taking the individual actions for an object $x$ and
$t\in T$ as follows:
\begin{eqnarray*}
R^{opt}(a_{P}|[x])&=&PP^{L}_{\eta}(t)P(X|[x])+PN^{L}_{\eta}(t)P(\neg
X|[x]);\\
R^{opt}(a_{B}|[x])&=&BP^{L}_{\eta}(t)P(X|[x])+BN^{L}_{\eta}(t)P(\neg
X|[x]);\\
R^{opt}(a_{N}|[x])&=&NP^{L}_{\eta}(t)P(X|[x])+NN^{L}_{\eta}(t)P(\neg
X|[x]).
\end{eqnarray*}

The bayesian decision procedure suggests the local minimum-cost
decision rules:

$(P):$ If $R^{opt}(a_{P}|[x])\leq R^{opt}(a_{B}|[x])$ and
$R^{opt}(a_{P}|[x])\leq R^{opt}(a_{N}|[x]),$ then $x\in POS(X);$

$(B):$ If $R^{opt}(a_{B}|[x])\leq R^{opt}(a_{P}|[x])$ and
$R^{opt}(a_{B}|[x])\leq R^{opt}(a_{N}|[x]),$ then $x\in BND(X);$

$(N):$ If $R^{opt}(a_{N}|[x])\leq R^{opt}(a_{P}|[x])$ and
$R^{opt}(a_{N}|[x])\leq R^{opt}(a_{B}|[x]),$ then $x\in NEG(X).$

Suppose $0\leq PP^{L}_{\eta}(t)\leq BP^{L}_{\eta}(t)\leq
NP^{L}_{\eta}(t)$ and $0\leq NN^{L}_{\eta}(t)\leq
BN^{L}_{\eta}(t)\leq PN^{L}_{\eta}(t)$ for $t\in T$. Since
$P(X|[x])+P(\neg X|[x])=1$, we simplify the rules as follows:

$(P):$ If $P(X|[x])\geq \alpha^{opt}(t)$, then  $x\in POS(X);$

$(B):$ If $\beta^{opt}(t)<P(X|[x])< \alpha^{opt}(t)$, then $x\in BND(X);$

$(N):$ If $P(X|[x])\leq \beta^{opt}(t)$, then $x\in NEG(X),$
\noindent where
\begin{eqnarray*}
\alpha^{opt}(t)=\frac{PN^{L}_{\eta}(t)-BN^{L}_{\eta}(t)}{PN^{L}_{\eta}(t)-BN^{L}_{\eta}(t)+BP^{L}_{\eta}(t)-PP^{L}_{\eta}(t)},
\beta^{opt}(t)=\frac{BN^{L}_{\eta}(t)-NN^{L}_{\eta}(t)}{BN^{L}_{\eta}(t)-NN^{L}_{\eta}(t)+NP^{L}_{\eta}(t)-BP^{L}_{\eta}(t)}.
\end{eqnarray*}
\begin{example} Let $\widetilde{\lambda}_{PP}(t),
\widetilde{\lambda}_{BP}(t), \widetilde{\lambda}_{NP}(t),
\widetilde{\lambda}_{NN}(t), \widetilde{\lambda}_{BN}(t)$ and $
\widetilde{\lambda}_{PN}(t)$ be shown as follows:
\begin{eqnarray*}
\widetilde{\lambda}_{PP}(t)&=&\frac{t}{1-\frac{1}{3t}}+\frac{t+1}{1-\frac{1}{3t}}
+\frac{2t+2}{1-\frac{1}{3t}}+\frac{3t+3}{1-\frac{1}{t}}+
\frac{5t+3}{1-\frac{1}{2t}}+\frac{4t+6}{1-\frac{1}{2t}}
+\frac{4t+8}{1-\frac{1}{2t}}+
\frac{4t+9}{1-\frac{1}{t}}+\frac{4t+10}{1-\frac{1}{t}},\\
\widetilde{\lambda}_{BP}(t)&=&\frac{2t+3}{1-\frac{1}{3t}}+\frac{2t+4}{1-\frac{1}{3t}}
+\frac{2t+5}{1-\frac{1}{3t}}+\frac{3t+6}{1-\frac{1}{t}}+
\frac{3t+7}{1-\frac{1}{2t}}+\frac{4t+6}{1-\frac{1}{2t}}
+\frac{4t+8}{1-\frac{1}{2t}}+
\frac{4t+9}{1-\frac{1}{t}}+\frac{4t+10}{1-\frac{1}{t}},
\end{eqnarray*}
\begin{eqnarray*}
\widetilde{\lambda}_{NP}(t)&=&\frac{3t+6}{1-\frac{1}{3t}}+\frac{3t+7}{1-\frac{1}{3t}}
+\frac{3t+8}{1-\frac{1}{3t}}+\frac{4t+9}{1-\frac{1}{t}}+
\frac{5t+9}{1-\frac{1}{2t}}+\frac{4t+10}{1-\frac{1}{2t}}
+\frac{4t+11}{1-\frac{1}{2t}}+
\frac{4t+12}{1-\frac{1}{t}}+\frac{5t+10}{1-\frac{1}{t}},\\
\widetilde{\lambda}_{NN}(t)&=&\frac{2t}{1-\frac{1}{3t}}+\frac{2t+1}{1-\frac{1}{3t}}
+\frac{2t+2}{1-\frac{1}{3t}}+\frac{3t+3}{1-\frac{1}{t}}+
\frac{5t+3}{1-\frac{1}{2t}}+\frac{4t+6}{1-\frac{1}{2t}}
+\frac{4t+8}{1-\frac{1}{2t}}+
\frac{4t+9}{1-\frac{1}{t}}+\frac{4t+10}{1-\frac{1}{t}},\\
\widetilde{\lambda}_{BN}(t)&=&\frac{3t+2}{1-\frac{1}{3t}}+\frac{3t+4}{1-\frac{1}{3t}}
+\frac{3t+6}{1-\frac{1}{3t}}+\frac{4t+6}{1-\frac{1}{t}}+
\frac{5t+6}{1-\frac{1}{2t}}+\frac{4t+7}{1-\frac{1}{2t}}
+\frac{4t+8}{1-\frac{1}{2t}}+
\frac{4t+9}{1-\frac{1}{t}}+\frac{4t+10}{1-\frac{1}{t}},
\\
\widetilde{\lambda}_{PN}(t)&=&\frac{4t+8}{1-\frac{1}{3t}}+\frac{4t+9}{1-\frac{1}{3t}}
+\frac{4t+10}{1-\frac{1}{3t}}+\frac{4t+11}{1-\frac{1}{t}}+
\frac{4t+12}{1-\frac{1}{2t}}+\frac{4t+13}{1-\frac{1}{2t}}
+\frac{5t+11}{1-\frac{1}{2t}}+
\frac{4t+15}{1-\frac{1}{t}}+\frac{5t+10}{1-\frac{1}{t}}.
\end{eqnarray*}

By taking $\eta=1-\frac{1}{3t}$, we have
\begin{eqnarray*}
\widetilde{\lambda^{\eta}}_{PP}(t)&=&[t,2t+2],
\widetilde{\lambda^{\eta}}_{PN}(t)=[4t+8,4t+10],
\widetilde{\lambda^{\eta}}_{BP}(t)=[2t+3,2t+5],\\
\widetilde{\lambda^{\eta}}_{BN}(t)&=&[3t+2,3t+6],
\widetilde{\lambda^{\eta}}_{NP}(t)=[3t+6,3t+8],
\widetilde{\lambda^{\eta}}_{NN}(t)=[2t,2t+2].
\end{eqnarray*}
Consequently, we have
\begin{eqnarray*}
\alpha^{opt}(t)&=&\frac{\lambda^{min}_{PN}(t)-\lambda^{min}_{BN}(t)}{\lambda^{min}_{PN}(t)-\lambda^{min}_{BN}(t)+\lambda^{min}_{BP}(t)-\lambda^{min}_{PP}(t)}=\frac{t+6}{2t+9};\\
\beta^{opt}(t)&=&\frac{\lambda^{min}_{BN}(t)-\lambda^{min}_{NN}(t)}{\lambda^{min}_{BN}(t)-\lambda^{min}_{NN}(t)+\lambda^{min}_{NP}(t)-\lambda^{min}_{BP}(t)}=\frac{t+2}{2t+5}.
\end{eqnarray*}
\end{example}

On the other hand,
$PP^{U}_{\eta}(t),BP^{U}_{\eta}(t),NP^{U}_{\eta}(t),PN^{U}_{\eta}(t)
,BN^{U}_{\eta}(t)$ and $NN^{U}_{\eta}(t)$ denote the upper bounds of
time-dependent fuzzy numbers in Table 4. We show the expected losses
$R^{pes}(a_{P}|[x])$, $R^{pes}(a_{B}|[x])$ and $R^{pes}(a_{N}|[x])$
associated with taking the individual actions for an object $x$ and
$t\in T$ as follows:
\begin{eqnarray*}
R^{pes}(a_{P}|[x])&=&PP^{L}_{\eta}(t)P(X|[x])+PN^{L}_{\eta}(t)P(\neg
X|[x]);\\
R^{pes}(a_{B}|[x])&=&BP^{L}_{\eta}(t)P(X|[x])+BN^{L}_{\eta}(t)P(\neg
X|[x]);\\
R^{pes}(a_{N}|[x])&=&NP^{L}_{\eta}(t)P(X|[x])+NN^{L}_{\eta}(t)P(\neg
X|[x]).
\end{eqnarray*}

The bayesian decision procedure suggests the local minimum-cost
decision rules:

$(P'):$ If $R^{pes}(a_{P}|[x])\leq
R^{pes}(a_{B}|[x]),R^{pes}(a_{P}|[x])\leq R^{pes}(a_{N}|[x])$, then
$x\in POS(X);$

$(B'):$ If $R^{pes}(a_{B}|[x])\leq
R^{pes}(a_{P}|[x]),R^{pes}(a_{B}|[x])\leq R^{pes}(a_{N}|[x])$, then
$x\in BND(X);$

$(N'):$ If $R^{pes}(a_{N}|[x])\leq
R^{pes}(a_{P}|[x]),R^{pes}(a_{N}|[x])\leq R^{pes}(a_{B}|[x])$, then
$x\in NEG(X).$

Suppose $0\leq PP^{U}_{\eta}(t)\leq BP^{U}_{\eta}(t)\leq
NP^{U}_{\eta}(t)$ and $0\leq NN^{U}_{\eta}(t)\leq
BN^{U}_{\eta}(t)\leq PN^{U}_{\eta}(t)$ for $t\in T$. Since
$P(X|[x])+P(\neg X|[x])=1$, we simplify the rules as follows:

$(P'):$ If $P(X|[x])\geq \alpha^{pes}(t)$, then  $x\in POS(X);$

$(B'):$ If $\beta^{pes}(t)<P(X|[x])< \alpha^{pes}(t)$, then $x\in BND(X);$

$(N'):$ If $P(X|[x])\leq \beta^{pes}(t)$, then $x\in NEG(X),$
\noindent where
\begin{eqnarray*}
\alpha^{pes}(t)=\frac{PN^{U}_{\eta}(t)-BN^{U}_{\eta}(t)}{PN^{U}_{\eta}(t)-BN^{U}_{\eta}(t)+BP^{U}_{\eta}(t)-PP^{U}_{\eta}(t)},
\beta^{pes}(t)=\frac{BN^{U}_{\eta}(t)-NN^{U}_{\eta}(t)}{BN^{U}_{\eta}(t)-NN^{U}_{\eta}(t)+NP^{U}_{\eta}(t)-BP^{U}_{\eta}(t)}.
\end{eqnarray*}
\begin{example}(Continuation of Example 5.5) On the basis of $\widetilde{\lambda}_{PP}(t),
\widetilde{\lambda}_{BP}(t), \widetilde{\lambda}_{NP}(t),
\widetilde{\lambda}_{NN}(t), \widetilde{\lambda}_{BN}(t)$ and $
\widetilde{\lambda}_{PN}(t)$ in Example 5.5, by taking
$\eta=1-\frac{1}{3t}$, we have
\begin{eqnarray*}
\alpha^{pes}(t)&=&\frac{\lambda^{max}_{PN}(t)-\lambda^{max}_{BN}(t)}{\lambda^{max}_{PN}(t)-\lambda^{max}_{BN}(t)+\lambda^{max}_{BP}(t)-\lambda^{max}_{PP}(t)}=\frac{t+4}{t+7};\\
\beta^{pes}(t)&=&\frac{\lambda^{max}_{BN}(t)-\lambda^{max}_{NN}(t)}{\lambda^{max}_{BN}(t)-\lambda^{max}_{NN}(t)+\lambda^{max}_{NP}(t)-\lambda^{max}_{BP}(t)}=\frac{t+4}{2t+7}.
\end{eqnarray*}
\end{example}

In general, by taking $PP_{\eta}(t)\in
\widetilde{\lambda^{\eta}}_{PP}(t),BP_{\eta}(t)\in
\widetilde{\lambda^{\eta}}_{BP}(t),NP_{\eta}(t)\in
\widetilde{\lambda^{\eta}}_{NP}(t),PN_{\eta}(t)\in
\widetilde{\lambda^{\eta}}_{PN}(t),BN_{\eta}(t)\in
\widetilde{\lambda^{\eta}}_{BN}(t)$ and $NN_{\eta}(t)\in
\widetilde{\lambda^{\eta}}_{NN}(t)$, we show the expected losses
$R(a_{P}|[x])$, $R(a_{B}|[x])$ and $R(a_{N}|[x])$ associated with
taking the individual actions for an object $x$ and $t\in T$ as
follows:
\begin{eqnarray*}
R(a_{P}|[x])&=&PP_{\eta}(t)P(X|[x])+PN_{\eta}(t)P(\neg
X|[x]);\\
R(a_{B}|[x])&=&BP_{\eta}(t)P(X|[x])+BN_{\eta}(t)P(\neg
X|[x]);\\
R(a_{N}|[x])&=&NP_{\eta}(t)P(X|[x])+NN_{\eta}(t)P(\neg X|[x]).
\end{eqnarray*}

The bayesian decision procedure suggests the local minimum-cost
decision rules:

$(P''):$ If $R(a_{P}|[x])\leq R(a_{B}|[x]),R(a_{P}|[x])\leq
R(a_{N}|[x])$, then $x\in POS(X);$

$(B''):$ If $R(a_{B}|[x])\leq R(a_{P}|[x]),R(a_{B}|[x])\leq
R(a_{N}|[x])$, then $x\in BND(X);$

$(N''):$ If $R(a_{N}|[x])\leq R(a_{P}|[x]),R(a_{N}|[x])\leq
R(a_{B}|[x])$, then $x\in NEG(X).$

Suppose $0\leq PP_{\eta}(t)\leq BP_{\eta}(t)\leq NP_{\eta}(t)$ and
$0\leq NN_{\eta}(t)\leq BN_{\eta}(t)\leq PN_{\eta}(t)$ for $t\in T$.
Since $P(X|[x])+P(\neg X|[x])=1$, we simplify the rules as follows:

$(P''):$ If $P(X|[x])\geq \alpha(t)$,
then  $x\in POS(X);$

$(B''):$ If $\beta(t)<P(X|[x])< \alpha(t)$, then
$x\in BND(X);$

$(N''):$ If $P(X|[x])\leq \beta(t)$,
then $x\in NEG(X),$
\noindent where
\begin{eqnarray*}
\alpha(t)=\frac{PN_{\eta}(t)-BN_{\eta}(t)}{PN_{\eta}(t)-BN_{\eta}(t)+BP_{\eta}(t)-PP_{\eta}(t)},
\beta(t)=\frac{BN_{\eta}(t)-NN_{\eta}(t)}{BN_{\eta}(t)-NN_{\eta}(t)+NP_{\eta}(t)-BP_{\eta}(t)}.
\end{eqnarray*}

On the basis of the above results, we have the following theorem for
DTRS when loss functions are time-dependent
fuzzy numbers.

\begin{theorem} Let $0\leq
PP^{L}_{\eta}(t)\leq PP^{U}_{\eta}(t)\leq BP^{L}_{\eta}(t) \leq
BP^{U}_{\eta}(t)\leq NP^{L}_{\eta}(t)\leq NP^{U}_{\eta}(t)$ and
$0\leq NN^{L}_{\eta}(t)\leq NN^{U}_{\eta}(t)\leq BN^{L}_{\eta}(t)
\leq BN^{U}_{\eta}(t)\leq PN^{L}_{\eta}(t)\leq PN^{U}_{\eta}(t)$,
where $t\in T$. Then
\begin{eqnarray*}
(1)
&&\alpha(t)\in[\frac{PN^{L}_{\eta}(t)-BN^{U}_{\eta}(t)}{PN^{U}_{\eta}(t)-BN^{L}_{\eta}(t)+BP^{U}_{\eta}(t)-PP^{L}_{\eta}(t)},
min\{\frac{PN^{U}_{\eta}(t)-BN^{L}_{\eta}(t)}
{PN^{L}_{\eta}(t)-BN^{U}_{\eta}(t)+BP^{L}_{\eta}(t)-PP^{U}_{\eta}(t)},1\}];\\
(2)&& \beta(t)\in[\frac{BN^{L}_{\eta}(t)-NN^{U}_{\eta}(t)}
{BN^{U}_{\eta}(t)-NN^{L}_{\eta}(t)+NP^{U}_{\eta}(t)-BP^{L}_{\eta}(t)},
min\{\frac{BN^{U}_{\eta}(t)-NN^{L}_{\eta}(t)}{BN^{L}_{\eta}(t)-NN^{U}_{\eta}(t)
+NP^{L}_{\eta}(t)-BP^{U}_{\eta}(t)},1\}].\end{eqnarray*}
\end{theorem}

\noindent\textbf{Proof.} $(1)$  Since $0\leq PP^{L}_{\eta}(t)\leq
PP^{U}_{\eta}(t)\leq BP^{L}_{\eta}(t) \leq BP^{U}_{\eta}(t)\leq
NP^{L}_{\eta}(t)\leq NP^{U}_{\eta}(t)$ and $0\leq
NN^{L}_{\eta}(t)\leq NN^{U}_{\eta}(t)\leq BN^{L}_{\eta}(t) \leq
BN^{U}_{\eta}(t)\leq PN^{L}_{\eta}(t)\leq PN^{U}_{\eta}(t)$, we have
$$PN_{\eta}(t)-BN_{\eta}(t)>0,BP_{\eta}(t)-PP_{\eta}(t)\geq 0.$$ It implies that
\begin{eqnarray*}
&&PN^{L}_{\eta}(t)-BN^{U}_{\eta}(t)\leq PN_{\eta}(t)-BN_{\eta}(t)
\leq
PN^{U}_{\eta}(t)-BN^{L}_{\eta}(t),\\&&BP^{L}_{\eta}(t)-PP^{U}_{\eta}(t)\leq
BP_{\eta}(t)-PP_{\eta}(t) \leq BP^{U}_{\eta}(t)-PP^{L}_{\eta}(t).
\end{eqnarray*} It follows that
\begin{eqnarray*}
PN^{L}_{\eta}(t)-BN^{U}_{\eta}(t)+ BP^{L}_{\eta}(t)-PP^{U}_{\eta}(t)
&\leq& PN_{\eta}(t)-BN_{\eta}(t)+ BP_{\eta}(t)-PP_{\eta}(t)\\&\leq&
PN^{U}_{\eta}(t)-BN^{L}_{\eta}(t)+
BP^{U}_{\eta}(t)-PP^{L}_{\eta}(t).
\end{eqnarray*} Obviously, we have
\begin{eqnarray*}
\frac{PN^{L}_{\eta}(t)-BN^{U}_{\eta}(t)}
{PN^{U}_{\eta}(t)-BN^{L}_{\eta}(t)+BP^{U}_{\eta}(t)-PP^{L}_{\eta}(t)}&\leq&
\frac{PN_{\eta}(t)-BN_{\eta}(t)}{PN_{\eta}(t)-BN_{\eta}(t)
+BP_{\eta}(t)-PP_{\eta}(t)}\\&\leq&
\frac{PN^{U}_{\eta}(t)-BN^{L}_{\eta}(t)}{PN^{L}_{\eta}(t)-BN^{U}_{\eta}(t)
+BP^{L}_{\eta}(t)-PP^{U}_{\eta}(t)}.
\end{eqnarray*}
Therefore,
$$\alpha(t)\in[\frac{PN^{L}_{\eta}(t)-BN^{U}_{\eta}(t)}{PN^{U}_{\eta}(t)-BN^{L}_{\eta}(t)+BP^{U}_{\eta}(t)-PP^{L}_{\eta}(t)},
min\{\frac{PN^{U}_{\eta}(t)-BN^{L}_{\eta}(t)}
{PN^{L}_{\eta}(t)-BN^{U}_{\eta}(t)+BP^{L}_{\eta}(t)-PP^{U}_{\eta}(t)},1\}].$$

$(2)$  Since $0\leq PP^{L}_{\eta}(t)\leq PP^{U}_{\eta}(t)\leq
BP^{L}_{\eta}(t) \leq BP^{U}_{\eta}(t)\leq NP^{L}_{\eta}(t)\leq
NP^{U}_{\eta}(t)$ and $0\leq NN^{L}_{\eta}(t)\leq
NN^{U}_{\eta}(t)\leq BN^{L}_{\eta}(t) \leq BN^{U}_{\eta}(t)\leq
PN^{L}_{\eta}(t)\leq PN^{U}_{\eta}(t)$, we have
$$BN_{\eta}(t)-NN_{\eta}(t)>0,NP_{\eta}(t)-BP_{\eta}(t)\geq 0 .$$ It
implies that
\begin{eqnarray*}
BN^{L}_{\eta}(t)-NN^{U}_{\eta}(t)&\leq& BN_{\eta}(t)-NN_{\eta}(t)
\leq
BN^{U}_{\eta}(t)-NN^{L}_{\eta}(t),\\NP^{L}_{\eta}(t)-BP^{U}_{\eta}(t)&\leq&
NP_{\eta}(t)-BP_{\eta}(t) \leq NP^{U}_{\eta}(t)-BP^{L}_{\eta}(t).
\end{eqnarray*} It follows that
\begin{eqnarray*}
BN^{L}_{\eta}(t)-NN^{U}_{\eta}(t)+ NP^{L}_{\eta}(t)-BP^{U}_{\eta}(t)
&\leq& BN_{\eta}(t)-NN_{\eta}(t)+ NP_{\eta}(t)-BP_{\eta}(t)\\&\leq&
BN^{U}_{\eta}(t)-NN^{L}_{\eta}(t)+
NP^{U}_{\eta}(t)-BP^{L}_{\eta}(t).
\end{eqnarray*} Obviously, we have
\begin{eqnarray*}
\frac{BN^{L}_{\eta}(t)-NN^{U}_{\eta}(t)}
{BN^{U}_{\eta}(t)-NN^{L}_{\eta}(t)+NP^{U}_{\eta}(t)-BP^{L}_{\eta}(t)}&\leq&
\frac{BN_{\eta}(t)-NN_{\eta}(t)}
{BN_{\eta}(t)-NN_{\eta}(t)+NP_{\eta}(t)-BP_{\eta}(t)}\\&\leq&
\frac{BN^{U}_{\eta}(t)-NN^{L}_{\eta}(t)}
{BN^{L}_{\eta}(t)-NN^{U}_{\eta}(t)+NP^{L}_{\eta}(t)-BP^{U}_{\eta}(t)}.
\end{eqnarray*}
Therefore,
$$\beta(t)\in[\frac{BN^{L}_{\eta}(t)-NN^{U}_{\eta}(t)}
{BN^{U}_{\eta}(t)-NN^{L}_{\eta}(t)+NP^{U}_{\eta}(t)-BP^{L}_{\eta}(t)},
min\{\frac{BN^{U}_{\eta}(t)-NN^{L}_{\eta}(t)}{BN^{L}_{\eta}(t)-NN^{U}_{\eta}(t)
+NP^{L}_{\eta}(t)-BP^{U}_{\eta}(t)},1\}].\Box$$

\section{Conclusions}

Many researchers have focused on investigations of loss functions in
DTRS. In this paper, we have investigated
DTRS when loss functions are satisfied time-dependent
uniform distributions and normal processes. Furthermore, we have
studied DTRS when loss functions are
time-dependent interval sets. Consequently, we have investigated DTRS when loss functions are time-dependent fuzzy
numbers. Finally, we have employed several examples to illustrate
that how to make decisions by using time-dependent loss functions-based
DTRS.

There are still many interesting topics deserving further
investigations on DTRS. For example, there are
many types of loss functions which are satisfied stochastic
processes, and it is of interest to investigate time-dependent loss
functions-based DTRS. In the future, we will
further investigate time-dependent loss functions and discuss the
application of DTRS in knowledge discovery.

\section*{ Acknowledgments}

We would like to thank the anonymous reviewers very much for their
professional comments and valuable suggestions. This work is
supported by the National Natural Science Foundation of China (NO.
11371130,11401052,11401195), the Scientific Research Fund of Hunan
Provincial Education Department(No.14C0049).

\end{document}